\documentclass{svjour2}                    % onecolumn
\smartqed  % flush right qed marks, e.g. at end of proof
\usepackage{graphicx}
\usepackage{amsmath}
\usepackage{amsfonts}
\usepackage{amssymb}

\begin{document}

\title{Two problems of theory of gravitation}


\author{L. Verozub%         \and
        %Second Author %etc.
}


\institute{L. Verozub \at
              Kharkov National University \\
             % Tel.: +123-45-678910\\
              \email{lverozub@gmail.com}           %  \\
        %   \and
          % S. Author \at
           %   second address
}

\date{Received: date / Accepted: date}

\maketitle

\begin{abstract}
This paper aims to discuss two issues that can have a significant impact on the foundations 
of the theory of gravitation:

\noindent
1. The existence of   relativity  of space-time geometry with respect to the  properties  of 
  used   reference frame, which is a manifestation of the  long-known fact of relativity of
 geometry of space and time with respect to    properties of 
 measuring instruments (Henri Poincare). 

\noindent
2. Lack of invariance of Einstein's equations with respect to the geodetic  transformations
 preserving unchanged the equations of motion of test particles. Because of this,
 the physically equivalent states  are described, generally speaking, by means of  
 different solutions of these equations.  In other words, there is no one-to-one 
correspondence between the solutions of these equations and the set of admissible 
physical states
%Include keywords, PACS and mathematical
%subject classification numbers as needed.
\end{abstract}

\section{Relativity of Space-Time}

Einstein's theory of gravity is a realization of the idea of the relativity of the 
properties of space-time with respect to the distribution of matter. However, 
it is well known that before the 
advent 
of Einstein's theory, Henri Poincar\'{e} showed that the properties of space and 
time are also 
relative to the properties of the used measuring instruments \cite{Poincare}.
Of course now it can be said also about the properties of space-time too. However, 
these convincing arguments have never
 been implemented in physical theory. 

We can make a step towards the realization of this idea, if we will pay attention 
 that the properties of measuring instruments are one of the characteristics of 
the used reference frame.
 We can, therefore, susapect that  we deal with the fact of
%the manifestation  of a fundamental property of physical reality --- with 
space-time relativity with respect to the used reference frame.

By a non-inertial frame of reference (NIRF) we mean the frame, the body of reference
of which is formed by point masses moving in an IFR under the effect of a
given force field. At that,  we  postulated, according to Special Relativity, that space-time
in inertial reference frame (IFR) is pseudo-Euclidean. On this basis one can find the 
line element of space-time  in the NIFR from the
viewpoint of observers located in the NIFR and proceeded from relativity of
space and time in the Bercley-Leibnitz-Mach-Poincar\'{e} (BLMP)  meaning.

The reference body (RB) of a reference frame is supposed to be formed by identical
 point masses $m$. If an observer in the reference frame   is at rest , his world
line coincides with the world line of some point of the reference body. It is
obvious for such an observer in an IFR that the accelerations of the point
masses forming his reference body are equal to zero.  This fact
takes place in relativistic meaning, too. That is, if the line element 
 of space-time in the IFR is denoted by $d\sigma$ and $u^{\alpha
}=dx^{\alpha}/d\sigma$ is the field  of 4- velocity of the
 point masses
forming the reference body, then the absolute derivative of 
$u^{\alpha}$ is equal to zero: \footnote{We use notations and
definitions, following the Landau and Lifshitz book \cite{Landau}.}
\begin{equation}
Du^{\alpha}/d\sigma=0. \label{DuDsigma}%
\end{equation}
(We mean that an arbitrary coordinate system is used.)

Does it occur for an observer in the NIFR ? That is, if the differential metric
form of space-time in a NIFR is denoted by $ds$, does the 4-velocity vector
$\zeta^{\alpha}=dx^{\alpha}/ds$ of the point-masses forming the reference body
of the NIFR satisfy the equation
\begin{equation}
D\zeta^{\alpha}/ds=0\, \, ?
\label{DzetaDs}%
\end{equation}
 If the space-time is absolute, equation (\ref{DzetaDs}) holds for only
$ds = d\sigma$. However, if space and time are relative in the BLMP sense, then
for both observers, located in some IFR and NIFR, the motion of the point masses,
forming their reference bodies, which are kinematically equivalent, must be
dynamically equivalent, too (both in non-relativistic and relativistic sense).
Any observer in the NIFR, isolated from the external world and proceeded from
relativity of space-time in BLMP meaning, consider points of the reference
body as the ones of his physical space, and space of events as his space-time.
Therefore, from his viewpoint point masses forming the reference body of his
frame are not under action of any forces (the same as for the observer in
IFR), and their 4-velocity must be equal to zero. In other words, since for
the observer in the IFR world lines of the reference body
are, according to  (\ref{DuDsigma}), some geodesic lines, for the observer
in the NIFR the world lines of the  of his BR also must be geodesic
lines in his space-time, which can be expressed by  (\ref{DzetaDs} ).

The equation (\ref{DzetaDs}) uniquely determines the fundamental metric form
in the above NIRF. Indeed, the differential equations of these world lines are at the
same time Lagrange equations describing in Minkowski space-time the motion of
the point masses forming the reference bodies of the NIFR. The last equations
can be obtained from a Lagrange action $S$ by the principle of the least
action. Therefore, the equations of the geodesic lines can be obtained from a
differential metric form $ds=k\ dS$, where $k$ is  constant, $dS=\mathcal{
L}(x,\dot{x})dt$, and $\mathcal{L}(x,\dot{x})$ is a Lagrange function
describing in Minkowski space-time the motion of identical point masses $m$
forming the body reference of the NIFR. . The constant $k$ is equal to
$-(mc)^{-1}$, as it follows from the analysis of the case when the frame of
reference is inertial.

The above NIFR can be named a proper reference frame (PRF) of the force field given in a IFR
Thus, if we proceed from relativity of space and time in the BLMP sense, then
the line element of space-time in PRFs can be expected to have
the following form:
\begin{equation}
ds=-(mc)^{-1}\;dS(x,dx). \label{dsMain}%
\end{equation}

Therefore, properties of space-time in PRFs are entirely determined by
properties of used frames in accordance with the BLMP idea of relativity of
space and time.

\section{Examples}

The above NIFR can be named a proper reference frame (PRF) of the force field given in a IFR. 
Consider  examples of PRFs.

1. The reference body is formed by noninteracting electric charges, moving in
a constant homogeneous electric field $\mathcal{E}$. The motion of the charges
in an IFR is described in the Cartesian coordinates system by a Lagrangian
\cite{Landau}
\begin{equation}
L=-m\,c^{2}\ (1-v^{2}/c^{2})^{1/2}+\mathcal{E}\,e\,x, \label{lagrframe1}%
\end{equation}
where $v$ is the speed of a particle. According to  (\ref{dsMain}) the
space-time metric differential form in this frame is given by%
\begin{equation}
ds=d\sigma-(wx/c^{2})dx^{0},
\end{equation}
where
\[
d\sigma=[\eta_{\alpha\beta}dx^{\alpha}dx^{\beta}]^{1/2}%
\]
is the differential metric form of the Minkowski space-time in the IFR in the
coordinate system being used, and $w=e\,\mathcal{E}/m$ is the acceleration of
the charges.

2. The reference body consists of noninteracting electric charges in a
constant homogeneous magnetic field $H$ directed along the axis $z$. The
Lagrangian describing the motion of the particles can be written as follows
\cite{Landau}:
\begin{equation}
L=-mc^{2}(1-v^{2}/c^{2})^{1/2}-(m\Omega_{0}/2)(\dot{x}y-x\dot{y}),
\label{LagrangGarge}%
\end{equation}
where $\dot{x}=dx/dt$, $\dot{y}=dx/dt$, and $\Omega_{0}=eH/2mc$.

The points of such a system rotate in the plane $xy$ around the axis $z$ with
the angular frequency
\begin{equation}
\omega=\Omega_{0}[1+(\Omega_{0}r/c)^{2}]^{-1/2} \label{omega (r)},%
\end{equation}
where $r=(x^{2}+y^{2})^{1/2}$. The linear velocity of the BR points tends to
$c$ when $r\rightarrow\infty$.

For the given NIFR
\begin{equation}
\label{dsInRotationNIFR}ds = d\sigma+ (\Omega_{0} / 2c)\ (ydx - xdy).
\end{equation}

Thus in the above examples  space-time is a Finslerian.

3. Another, rather unexpected example, give the recent results on the motion of small
elements of a perfect isentropic fluid \cite{Verozub08a}.

Instead of the traditional continuum assumption, the behavior of the fluid flow can 
be considered as the motion of a finite numbler of particles under the influence of interparticles
 forces which mimic effects of pressure, viscosity, etc. \cite{Monaghan}. 
Owing to replacement of integration by summation over a number of particles,   continual
 derivatives  become simply  time derivatives along  the particles trajectories. 
The velocity of the fluid at a given point is the velocity of the particle at this point. 
The continuity equation is always fulfilled and can consequently be omitted.  
Owing to such discretization the motion of particles is governed by means of solutions of 
ordinary differential equations of classical or relativistic dynamics.

In \cite{Verozub08a} it was shown  
 that the  Lagrangian described the motion  of  macroscopically small elements  (``particles'') of a perfect isentropic fluid    is given by
\begin{equation}
L=-mc \left( G_{\alpha \beta }\frac{dx^{\alpha }}{d\lambda }\frac{%
dx^{\beta }}{d\lambda }\right) ^{1/2}d\lambda   \label{Lagrangian_in_V},
\end{equation}%
In this equation 
$m$ are the masses of the ``particles'',  $c$ is speed of light,
 $G_{\alpha \beta }=\varkappa ^{2}\, \eta_{\alpha \beta }$, where
 $\eta_{\alpha \beta}$ is the metric tensor of the  Minkowski space-time $E$, 
\begin{equation}
\varkappa =\frac{w}{nmc^{2}}=1+\frac{\varepsilon }{\rho c^{2}}+\frac{P}{\rho
c^{2}},  \label{xi},
\end{equation}%
where  $w$ is the fluid enthalpy per unit of volume, 
$\varepsilon $ is the fluid density energy, 
 $\rho=m n$, $n$ is the particles number density, $P$ is the 
pressure i the fluid,
$\lambda $ is a parameter along  4-paths of particles.

In an inertial reference frame (i.e. in Minkowski space-time $E$)
 we can set the parameter $\lambda =\sigma $
which yields the following  Lagrange equations which does not contain the mass $m$:

\begin{equation}
\frac{d}{d\sigma }\left( \varkappa u_{\alpha }\right) -\frac{\partial
\varkappa }{\partial x^{\alpha }}=0
\end{equation}%
where $u_{\alpha }=\eta _{\alpha \beta }u^{\beta }$, and $u^{\alpha}=dx^{\alpha}/d\sigma$. 
 For adiabatic processes 
\cite{Landau} 
\begin{equation}
\frac{\partial }{\partial x^{\alpha}}\left( \frac{w }{n}\right) =\frac{1}{n}\frac{\partial P%
}{\partial x^{\alpha }},
\end{equation}%
and we arrive at the equations of the motion of the set of the particles in the form%
\begin{equation}
w\frac{du_{\alpha }}{d\sigma }+u_{\alpha }u^{\beta }\frac{\partial P}{%
\partial x^{\beta }}-\frac{\partial P}{\partial x^{\alpha }}=0.
\label{MotionEquation_in_E}
\end{equation}%
where $du_{\alpha }/d\sigma =\left( \partial u_{\alpha }/\partial
x^{\epsilon }\right) u^{\epsilon }.$ It is the general accepted relativistic
equations of the motion of fluid  \cite{Landau}.

In a comoving reference frame the space-time the line element is of the form
\begin{equation}
ds^{2}=G_{\alpha \beta }dx^{\alpha }dx^{\beta }  \label{ds2}.
\end{equation}%

In this case the element of the proper time is $ds$. After the setting 
$\lambda =s$,
the Lagrangian equation of the motion takes the standard form of a congruence of 
geodesic lines :%
\begin{equation}
\frac{du^{\alpha }}{ds}+\Gamma _{\beta \gamma }^{\alpha }u^{\beta }u^{\gamma
}=0,  \label{eqsPart_as_Geodesic}
\end{equation}%
where $du_{\alpha }/ds=\left( \partial u_{\alpha }/\partial x^{\epsilon
}\right) u^{\epsilon }$, $u^{\alpha}=du^{\alpha}/ds$,
and 
\begin{equation}
\Gamma _{\beta \gamma }^{\alpha } =\frac{1}{2}G^{\alpha \epsilon }\left( 
\frac{\partial G_{\epsilon \beta }}{\partial x^{\gamma }}+\frac{\partial
G_{\epsilon \gamma }}{\partial x^{\beta }}-\frac{\partial G_{\beta \gamma }}
{\partial x^{\epsilon }}\right).
\end{equation}
In the Cartesian coordinates
\begin{equation}
\Gamma_{\alpha\beta}^{\gamma}=
\frac{1}{\varkappa }\left( \frac{\partial \varkappa }{\partial x^{\gamma }%
}\delta _{\beta }^{\alpha }+\frac{\partial \varkappa }{\partial x^{\beta }}%
\delta _{\gamma }^{\alpha }-\eta ^{\alpha \epsilon }\frac{\partial \varkappa 
}{\partial x^{\epsilon }}\eta _{\beta \gamma }\right),
\end{equation}
so that 
\begin{equation}
 \Gamma_{00}^{1}= -\frac{1}{\rho c^{2}}\frac{\partial P}{\partial x^{1}}
\end{equation}

In the spherical coordinates the scalar curvature $R$ is given by
\begin{equation}
 R = \frac{6}{\varkappa^{3} r^{2}} (r^{2} \varkappa')', 
\end{equation}
where the prime denotes a derivative with respect to $r$.

Therefore, the motion of small elements of the fluid in a comoving reference 
frame can be viewed as the motion in a Riemannian space-time with a nonzero 
curvature.

 4. Suppose that in the Minkowski space-time gravitation can be described
as a tensor field  $\psi_{\alpha\beta}(x)$ in $E$, and the Lagrangian,
describing the motion of a test particle with the mass $m$ in $E$  is of
the form
\begin{equation}
\mathcal{L}=-m c [g_{\alpha\beta}(\psi)\;\dot{x}^{\alpha}\;\dot
{x}^{\beta}]^{1/2},  \label{LagrangianThirr}
\end{equation}
where $\dot{x}^{\alpha}=dx^{\alpha}/dt$ and $g_{\alpha\beta}$ is a
symmetric tensor whose components are 
functions of $ \psi_{\alpha\beta} $  \cite{Thirring}.
If particles move under influence of the force field $\psi_{\alpha\beta}(x)$,  
then according to (\ref{dsMain})
 the space-time line element in PFRs of this field takes the form
\begin{equation}
ds^{2}=g_{\alpha\beta}(\psi)\;dx^{\alpha}\;dx^{\beta}
\end{equation}
Consequently,  the space-time in
such PRFs is  Riemannian  $V$  with  curvature other than zero.
The tensor $g_{\alpha\beta}(\psi)$  is a  space-time metric tensor in  the PRFs. 

%Evidently, in this case we deal with Einstein's gravity.

Viewed by an
observer located in the IRF, the motion of the particles, forming the
reference body of the PRF, is affected by the force field $\psi_{\alpha\beta} $. 
Let $x^{i}(t,\chi)$ be a set of the particles paths, depending on the parameter $\chi$.
 Then, for the observer located in the IRF the
relative motion of a pair of particles from the set is described in
non-relativistic limit by the differential equations \cite{MTU} 
\begin{equation}
\frac{\partial^{2}n^{i}}{\partial t^{2}}+\frac{\partial^{2}U}{\partial
x^{i}\partial x^{k}}n^{k}=0,  \label{DevIFR}
\end{equation}
where $n^{k}=\partial x^{k}/\partial{\chi}$ and $U$ is the gravitational
potential.

However, the observer in a PRF of this field will not feel the existence of
the field.% since he moves in the space-time of the PRF along a geodesic line.
The presence of the field $\psi_{\alpha\beta}$ will be displayed for him
differently --- as a space-time curvature which manifests itself as a deviation
of the world lines of nearby points of the reference body.

For a quantitative description of this fact it is natural for him to use the
Riemannian normal coordinates.
\footnote{This and the above consideration does not depend on the used coordinate 
system, it can be performed by a covariant method.}
 In these coordinates spatial components of
the deviation equations of geodesic lines are 
\begin{equation}
\frac{\partial n^{i}}{\partial t^{2}}+R_{0k0}^{i}n^{k}=0,
\end{equation}
where $R_{0k0}^{i}$ are the components of the Riemann tensor. 
In the Newtonian limit these equations coincide with (\ref{DevIFR}).

Thus, in two above frames of reference  we have two different descriptions of 
particles 
motion --- as moving under the action of a force field in the Mankowski space-time, and as 
moving 
along the geodesic line in a Riemann space-time
with the curvature other than zero.

Of course,  (\ref{dsMain}) refers to any classical field $\mathcal{F}$. In particular, 
space-time in PRFs of an electromagnetic field is Finslerian \cite{Verozub06}. However, since 
$ds$,  in this case, depends on the mass and charge  of the particles
forming the reference body, this fact is not of great significance.

Thus any force field  can be considered based on 
 the aggregate  ``IRF + Minkowski space'', and based on  the  aggregate 
 ``PRF + non-Euclidean space-time with metric (\ref{dsMain})``. 
From this point of view of  geometrization of gravity is the second possibility, 
which was discovered by Einstein's intuition.

\textit{It is important to realize that the relativity of space-time geometry to the 
frame of reference is the same important and fundamental property of physical relativity 
as relativity to  act of measurement, the physical 
realization of which is quantum mechanics}. Full implementation of these ideas 
can have far-reaching implications for fundamental physics.

%Correct equations of gravity, taking into account this circumstance, will make it 
%possible to unify gravity with other fields. 
%Our results \cite{Verozub1991}
%indicates that from such point of view the original Einstein's equations cannot be 
%true completely. 

\section{Gravity equations and gauge-invariance}
In the theory of gravitation  the equations of motion of test particles play a fundamental 
role.
 Notion of "gravitational field" emerged as something necessary to correctly describe the 
motion of bodies. The magnitudes  that appear in the equations of motion, 
become the main characteristic of the field. The field equations have emerged as a tool for 
finding these magnitudes for a given  distribution of masses.

All this is very similar to
 classical electrodynamics. In this case  the equations of motion of
 test charges are invariant under  gauge transformations of 4-potentials. For this reason, 
all  4-potentials, obtained from a given by a gauge transformation, describe the same 
field.
 That is why the field equations of classical electrodynamics are invariant under gauge 
transformations.

Einstein's  equations of the motions of test particles in gravitational field 
are also invariant with respect to some class of transformations of the field 
variables in  any given coordinate system  --- 
with respect to geodesic transformations of  Christoffel symbols (or metric tensor)
\cite{Eisenhart}.
 Such transformations for 
 the Christoffel symbols are of the form
\begin{equation}
\label{GammaGeodesTransformations}\overline{\Gamma}_{\beta\gamma}^{\alpha
}(x)=\Gamma_{\beta\gamma}^{\alpha}(x)+\delta_{\beta}^{\alpha}\ \phi_{\gamma
}(x)+\delta_{\gamma}^{\alpha} \phi_{\beta}(x),
\end{equation}
where $\phi_{\alpha}(x)$ is  a continuously differentiable vector
field.
(The transformations for the metric tensor are solutions of some complicate 
partial differential equations).

Consequently, all  Christoffel symbols obtained from a given by geodesic 
transformations,
 describe the same gravitational field. 
 The equations for determining the gravitational field must be invariant under 
such 
transformations, and the physical meaning can only have values which are invariant under
 geodesic transformations.

However, Einstein's gravitational equations are not consistent completely
with the requirement 
which imposes on them the  hypothesis of the motion of test particles along geodesics,
 because they are not geodesically invariant \cite{Petrov}. 

Therefore, we can assume that in a fully correct theory of gravity, based
 on the hypothesis 
of the motion of test particles along geodesics, geodesic transformations should play the 
role 
of gauge transformations, and coordinate transformations should play the same role as in 
electrodynamics.

Einstein equations are in good agreement with observations in weak and moderately strong 
fields.
Therefore, if there are more correct equation of gravitation, then deriving from them  
physical results should differ  observably from   Einstein's equations only in strong fields. 

Simplest vacuum equation of this kind 
 were first proposed (from a different point of view) in \cite{Verozub91}, 
and discussed in greater detail in \cite{Verozub06},
  their physical implications discussed in 
\cite{Verozub96} - \cite{Verozub06}, 
and the equations in the presence of matter - in \cite{VerKoch00}. 
They are some  geodesic-invariant modification of Einstein's equations.

From a theoretical point of view, the most satisfactory are the vacuum equations. 

They predict some fundamentally new physical consequences which can be tested 
experimentally.

%Consider which form should be the simpleast generalization of Einstein's
%vacuum equations.

Under geodesic transformations the Ricci tensor $R_{\alpha\beta}$ of space-time $V$  
in PRFs of gravitational field transforms 
as follows:
\begin{equation}
\overline{R}_{\alpha\beta}=R_{\alpha\beta}+(n-1)\psi_{\alpha\beta},
\end{equation}
where
\begin{equation}
\psi_{\alpha\beta}=\psi_{\alpha;\beta}-\psi_{\alpha}\psi_{\beta,}
\end{equation}
 and a semicolon denotes a covariant differentiation 
in $V$. Therefore, the simpest generalization of the Einstein
equations is of the form
\begin{equation}
R_{\alpha\beta}+(n-1)\Gamma_{\alpha\beta}=0,
\end{equation}
 where $\Gamma_{\alpha\beta}$ is a tensor transformed under geodesic
transformations as follows
\begin{equation}
\overline{\Gamma}_{\alpha\beta}=\Gamma_{\alpha\beta}-\psi_{\alpha\beta}.
\end{equation}

Due to the fact that our space-time is a bimetric, there exists a
vector field
\begin{equation}
Q_{\alpha}=\Gamma_{\alpha}-\overset{\circ}{\Gamma}_{\alpha}%=\frac{1}{2(n+1)}
%\frac{\partial}{\partial x^{\alpha}}log\left\Vert \frac{g}{\eta}\right\Vert ,
\end{equation}
where $\Gamma_{\alpha}=\Gamma_{\alpha\beta}^{\beta}$ , $\overset{\circ}{\Gamma}_{\alpha}=
\overset{\circ}{\Gamma}_{\alpha\beta}^{\beta}$
,  $\Gamma_{\alpha\beta}^{\gamma}$ and 
$\overset{\circ}{\Gamma}_{\alpha\beta}^{\gamma}$
 are the Christoffel symbols  
in $V$ and $E$, respectively.

Under geodesic transformations in $V$  the quantities $\Gamma_{\alpha}$ are transformed 
as follows:

\begin{equation}
 \overline{\Gamma}_{\alpha} =\Gamma_{\alpha} + (n+1)\,\psi_{\alpha}
\end{equation}
For this reason, a tensor object 
\begin{equation}
 A_{\alpha\beta}=Q_{\alpha;\beta}-Q_{\alpha}Q_{\beta},
\end{equation}
 where $Q_{\alpha;\beta}$ is a covariant derivative of $Q_{\alpha}$ in $V$, has the 
same transformation properties under geodesic transformations as  must have the above 
vector field $\Gamma_{\alpha\beta}$.

The line element of space-time in PRFs was obtained from 
the Lagrangian motion of test particles in the Minkowski space-time $E$. 
If we want to find the equation of gravity in space-time $E$, you must realize that
 in this space, 
the Christoffel symbols $\Gamma_{\alpha\beta}^{\gamma}$ 
can be regarded as components of a tensor $\Gamma_{\alpha\beta}^{\gamma}-
\overset{\circ}{\Gamma}_{\alpha\beta}^{\gamma}$ in the Cartesian coordinate system,
 i.e.  
as components of $\Gamma_{\alpha\beta}^{\gamma}$, where the ordinary derivatives 
replaced by covariant in the metric of space-time $E$.
(Just as in bimetric Rosen's theory  \cite{Treder}). 

Given this, we arrive at the conclusion that the equations
\begin{equation}
 R_{\alpha\beta} - A_{\alpha\beta}=0
\end{equation}
are  a  simplest geodesic invariant modification of the vacuum Einstein equations, considered 
from the point of view of flat space-time.

These equations can be written in another form.
The simplest geodesic-invariant object in $V$ is a Thomas symbols:
\begin{equation}
 \Pi_{\alpha\beta}^{\gamma}=\Gamma_{\alpha\beta}^{\gamma}-
\frac{1}{n+1}
\left(\delta_{\alpha}^{\eta} \Gamma_{\beta}+ 
      \delta_{\beta}^{\eta} \Gamma_{\alpha} \right). 
\end{equation}
It is  not a tensor. However, from point of view of flat space-time $E$, 
they can be considered as components of the tensor $B_{\alpha\beta}^{\gamma} =
\Pi_{\alpha\beta}^{\gamma}-\overset{\circ}{\Pi}_{\alpha\beta}^{\gamma}$,
where $\overset{\circ}{\Pi}_{\alpha\beta}^{\gamma}$ is the Thomas symbols in $E$.
In another words, $B_{\alpha\beta}^{\gamma}$ can  
be considered  as the Thomas symbols where  derivatives
replaced by the covariant ones with respect to the metric $\eta_{\alpha\beta}$.
This geodesic-invariant tensor can be named by strength tensor of gravitational field.

The above gravitation equation can be written by tensor $B_{\alpha\beta}^{\gamma}$ 
as follows:
\begin{equation}
\label{myequations}
\bigtriangledown_{\gamma} B_{\alpha\beta}^{\gamma}-
B_{\alpha\delta}^{\gamma} B_{\beta\gamma}^{\delta}=0.
\end{equation}
where $\bigtriangledown$ denotes  a covariant derivative in $E$.

The physical consequences following from these equations 
 do not contradict any observational data, however, lead to some unexpected results, 
which allow to us to test the theory.
The first result is that they predict the existence of supermassive compact objects 
without event horizon which are an alternative to supermassive black holes in the centers 
of galaxies.\cite{Verozub06}
 The second result  is that they provide a simple and natural explanation for the 
fact of an acceleration of the universe as of a consequence of the gravity properties \cite{Verozub08}.

\section{Remarks on the equations inside matter}

 We can not claim that the particles inside any material medium move along geodesics. 
%The exception is the case of dust matter and the perfect fluid. 
Consequently,  it is unclear whether the field equations inside the matter
 to be a generalization of the geodesic equations of Einstein. 
However, such equations have been proposed in the work \cite{VerKoch00}. 
Comparison of the results obtained from 
them with observations of the binary pulsar PSR 1913+16 shows good agreement with 
observations. 
Despite this,  doubts   as to their correctness are still remain.
The problem is that the writing of 
generalization of the  equations in the matter requires 
significantly 
narrow the class of admissible geodesic transformations of the metric tensor of 
space-time $V$. It is not clear  whether such space-time is Riemannian.
It is possible,  geodesic invariance is violated in a material medium.
For this reason,  we do not consider these equations here in 
more detail, assuming that this is still a subject for further research.

\end{document}